\documentclass[paper,traditabstract]{aa}
\usepackage{txfonts}

\usepackage{graphicx,longtable,lscape,stfloats,natbib,psfig,amssymb}

\usepackage{natbib}
\bibpunct{(}{)}{;}{a}{}{,} 
\newcommand{\ltsima} {$\; \buildrel < \over \sim \;$}  
\newcommand{\gtsima} {$\; \buildrel > \over \sim \;$}  
\newcommand{\lta} {\lower.5ex\hbox{\ltsima}}  
\newcommand{\gta} {\lower.5ex\hbox{\gtsima}}  
\newcommand{\Ha} {H$\alpha$}  
\newcommand{\Hb} {H$\beta$}  

\newcommand{\ergs}{\>{\rm erg}\,{\rm s}^{-1}}
\newcommand{\ergscm}{\>{\rm erg}\,{\rm s}^{-1}\,{\rm cm}^{-2}}

\newcommand{\oiii}{\forb{O}{III}}

\newcommand{\forb}[2]{\mbox{$[{\rm #1\, #2}]$}}
\newcommand{\ew}{EW{\tiny{[O~III]}}}
\newcommand{\tsp}{\,\,\,\,\,}
\newcommand{\loiii}{L$_{\rm{\tiny{ [O~III]}}}$}
\newcommand{\lx}{L$_{\rm{\tiny{X}}}$}
\begin{document}
\title{Multi-band constraints on the nature of emission line galaxies}
\subtitle{} \titlerunning{Multi-band constraints on the nature of emission
  line galaxies} \authorrunning{Balmaverde \& Capetti}

\author{
B.~Balmaverde\inst{1}
\and A.~Capetti\inst{1}}

\institute {INAF - Osservatorio Astrofisico di Torino, Via Osservatorio 20,
  I-10025 Pino Torinese, Italy}

\offprints{balmaverde@oato.inaf.it} 

\abstract {Our aim is to  explore the nature of emission line galaxies
by  combining  high-resolution  observations obtained  in  different
bands  to understand which  objects are  powered by  Active Galactic
Nucleus (AGN). From the spectroscopic Palomar survey of nearby bright
galaxies,  we selected  a sample  of 18  objects observed  with HST,
Chandra, and VLA.

No connection is found  between X-ray and emission line luminosities
from  ground-based   data,  unlike   what  is  found   for  brighter
AGN.  Conversely, a strong  correlation emerges  when using  the HST
spectroscopic  data,   which  are   extracted  on  a   much  smaller
aperture. This  suggests that  the HST data  better isolate  the AGN
component when one is  present, while ground-based line measurements
are affected by diffuse emission from the host galaxies.

The  sample  separates  into   two  populations.  The  11  objects
belonging to the first class  have an equivalent width (EW) of the
\oiii\  emission line  measured  from HST  data  \ew\ $\gtrsim$  2
\AA\  and are  associated with  an  X-ray nuclear  source; in  the
second  group  we  find  seven  galaxies  with  \ew  $\lesssim$  1
\AA\ that generally do not  show any emission related to an active
nucleus  (emission lines,  X-ray, or  radio sources).  This latter
group  includes   about  half   of  the  Low   Ionization  Nuclear
Emission-line  region  (LINERs)  or  transition  galaxies  of  the
sample, all  of which  are objects of  low \oiii\  line luminosity
($\lesssim  10^{38}\ergs$) and low  equivalent width  ($\lesssim 1
$\AA) in ground-based  observations.  These results strengthen the
suggestion that the \ew\ value is a robust predictor of the nature
of an emission line galaxy.}


\maketitle

\section{Introduction}

Emission  lines are  among the  most widely  used tools  to  reveal an
active  galactic nucleus  (AGN)  and to  explore  its properties.  The
luminosity of narrow  emission lines is a robust  estimator of the AGN
bolometric luminosity \citep{mulchaey96}  and emission line ratios can
distinguish  H~II  regions  from   gas  ionized  by  nuclear  activity
\citep{heckman80}  and  separate AGN  into  various subclasses,  e.g.,
Seyferts and LINERs \citep{kewley06}.

However,  the nature of  emission line  galaxies, and  particularly of
LINERs, is  controversial, and it  is still unclear which  objects are
indeed powered by an active nucleus  and how they can be isolated from
the  overall population.  We recently  showed that  the  host galaxies
produce  a  substantial  contamination   of  the  emission  lines.  In
\citet{capetti11c} we considered the spectra of $\sim$300,000 galaxies
with $z<0.1$ from  the Sloan Digital Sky Survey  (SDSS). We found that
the equivalent width distribution of the [O~III]$\lambda5007$ emission
line,  \ew,\footnote{\citet{ho97} and  \citet{shields07} quote  the EW
  only for  \Ha; we used these  data to measure the  ratio between the
  \oiii\   and   the   continuum   around   the   \Ha\   line,   i.e.,
  EW{\tiny{[O~III]@\Ha}                     =                    ${\rm
      log}\,F_{\rm[O~III]}/F_{\rm{cont,H_\alpha}}$}}    is    strongly
clustered around $\sim$0.6 \AA.  The same conclusion was obtained from
studying the  486 nearby  (average distance of  $\sim 15$  Mpc) bright
galaxies from  the Palomar survey  \citep{filippenko85,ho97}. The bulk
of these low  EW objects is formed by LINERs, but  they also include a
significant number of galaxies classified as Seyfert, although located
not far from the LINERs/Seyfert  dividing line. These results are very
difficult to account  for if the emission lines are  powered by an AGN
because it requires a fine-tuning  between the strength of the nuclear
ionizing field,  the spatial distribution  of the emission  lines, and
the  stellar  mass. Conversely,  the  strong  connection between  line
emission  and stellar  continuum points  to  a stellar  origin of  the
emission lines. In particular, it  has been suggested that hot evolved
stars       can      play       a      dominant       role      (e.g.,
\citealt{trinchieri91,binette94}),  an  idea  also  supported  by  the
results  obtained  from  photoionization   models  that  are  able  to
reproduce   the   observed  EW   and   optical   line  ratios   (e.g.,
\citealt{stasinska08,sarzi10}).  These  results  cast  doubts  on  the
reliability of  the identification of active galaxies  based solely on
optical  spectroscopy,  particularly  those  of  LINERs  of  low  line
luminosity.

The  signature  of an  active  nucleus can  also  be  sought in  other
observing  bands such  as in  the X-ray  and in  high-resolution radio
observations. Furthermore,  because the dominant  contamination of the
emission  lines  is  apparently  associated  with  stellar  processes,
significant progress can be made  by considering spectra obtained in a
smaller physical region, such as those produced with HST observations.

Our aim is  to use multiband data to obtain  a better understanding of
the nature of emission line  galaxies, i.e., which of these objects is
associated  with an  AGN, particularly  at low  luminosities.  Several
studies   were  already   performed  along   this  line   (see,  e.g.,
\citealt{nagar05,zhang09,gonzalez09}). The  element of novelty  of our
study is the inclusion of the \oiii\ equivalent width in the analysis,
which,  as  explained  above,  is   related  to  the  effects  of  the
contamination from  the host galaxies  to the emission lines.  We show
that as  a result of this  approach, a significant  improvement in the
classification of galaxies  in the various classes of  activity can be
achieved once the proper set of observational data is available.

\begin{table*}
\caption{Parameters of the sample galaxies.}
\begin{tabular}{|l| c|c l| c| c | c c  c c|  c|}
\hline
Name    & Distance & \multicolumn{2}{|c|}{Spectr. class} & \multicolumn{2}{|c|}{[O~III] luminosity (and EW)} &\multicolumn{4}{|c|}{Chandra observations} & L$_{\rm radio}$      \\ 
        &          & Ho & K06       &Palomar& HST & Obs. Id   & Exp. Time     & Cts        & L (2-10 keV)&    \\ 
\hline                                                                                         
NGC~2787 & 13.0 & L   & L  &  38.37  \tsp (1.1)    &    37.75 \tsp (   9.8) &   4689 & 30.8 & 476     &    39.11   &  \tsp 20.15 	  \\
NGC~3368 &  8.1 & L   & L  &  37.64  \tsp (0.6)    & $<$36.40 \tsp ($<$0.5) &   391  & 2.0  &  7      & $<$38.26   &   $<$18.89  \\
NGC~3489 &  6.4 & T/S & A (L/S) &  38.33  \tsp (1.9)    & $<$36.28 \tsp ($<$0.4) &   392  & 1.7  &  10     &    38.14   &   $<$18.69  \\
NGC~3982 & 17.0 & S   & S &  39.83  \tsp (76.1)   &    39.24 \tsp (   640) &   4845 & 9.2  &  52     &    38.69   &   $<$19.54  \\
NGC~3992 & 17.0 & T   & A (L/S) &  38.40  \tsp (1.3)    &    37.10 \tsp (   2.5) &   \multicolumn{4}{|c|}{No Chandra observations} &   $<$19.64  \\
NGC~4143 & 17.0 & L   & L &  38.81  \tsp (2.0)    &    38.49 \tsp (   25 ) &  1617  & 2.51 & 144     &    40.00   &   \tsp 20.06 	  \\
NGC~4203 &  9.7 & L   & L &  38.53  \tsp (2.4)    &    38.28 \tsp (   38 ) &  10535 & 41.6 &6144     &    40.14   &   \tsp 20.03 	  \\    
NGC~4314 &  9.7 & L   & A (SF/L) &  37.75  \tsp (0.6)    & $<$36.17 \tsp ($<$1.0) &  2062  & 16.1 &  20     & $<$37.78   &   $<$19.05  \\
NGC~4321 & 16.8 & T   & A (SF/L) &  38.24  \tsp (1.4)    &    37.20 \tsp (   2.4) &  6727  & 37.9 &  89     &    38.22   &   $<$19.48  \\
NGC~4435 & 16.8 & T/H & A (SF/L)  &  38.02  \tsp (0.3)    & $<$36.67 \tsp ($<$1.0) &  8042  & 4.90 &  10     & $<$38.71   &   $<$19.57  \\
NGC~4450 & 16.8 & L   & L &  38.78  \tsp (2.4)    &    38.43 \tsp (    24) &  3997  & 3.35 & 447     &    40.14   &  \tsp 19.83 	   \\
NGC~4459 & 16.8 & T   & A (SF/L) &  37.83  \tsp (0.2)    &    36.96 \tsp (   0.6) &  11784 & 29.8 &  71     & $<$38.75   &  $<$19.53   \\
NGC~4477 & 16.8 & S   & A (L/S) &  38.82  \tsp (2.7)    &    37.44 \tsp (   4.2) &  9527  & 37.7 &  81     &    38.62   &  $<$19.53   \\
NGC~4501 & 16.8 & S   & S &  39.10  \tsp (5.9)    &    37.74 \tsp (   6.1) &  2922  & 17.87& 188     &    39.39   &  $<$19.57   \\
NGC~4548 & 16.8 & L   & L &  38.11  \tsp (0.9)    &    37.10 \tsp (   2.5) &  1620  & 2.7  &  20     &    38.85   &   \tsp 19.61 	  \\
NGC~4596 & 16.8 & L   & L & $<$37.62\tsp ($<$0.3) & $<$36.70 \tsp ($<$0.4) &  11785 & 31.0 &  28     & $<$38.38   &  $<$19.57   \\
NGC~4698 & 16.8 & S   & A (L/S) &  38.81  \tsp (3.1)    &    37.70 \tsp (   8.8) &  3008  & 145  &  20     &    38.75   &  $<$19.53   \\
NGC~5055 &  7.2 & T   & L &  37.43  \tsp (0.4)    & $<$36.67 \tsp ($<$0.6) &  2197  & 28.0 & 209     &    38.54   &   $<$18.83  \\
\hline                                                                                         
\end{tabular}
\label{table}
\medskip

Column description: 1) Name, 2) distance in Mpc, 
3) spectroscopic classification according to \citet{ho97} (SF = star forming, L = LINER, S
= Seyfert, T = transition galaxy) and 4) adopting to the scheme of
\citet{kewley06} that also includes ambiguous galaxies (A), 5) logarithm of the \oiii\ luminosity ($\ergs$), and 6) equivalent 
width from the Palomar survey \citep{ho97}; 7 and 8) \oiii\ luminosity and EW from
the HST survey \citep{shields07}; data from the Chandra
observations: 9) observations ID, 10) exposure time (ks), 11) nuclear counts, 12) luminosity in the 2-10 keV range ($\ergs$); 
13) radio luminosity at 15 GHz (W/Hz) from \citet{nagar05}.
\end{table*}

\section{Sample and the multi-band data}
\label{sample}
\begin{figure*}
\centering
\includegraphics[width=17cm]{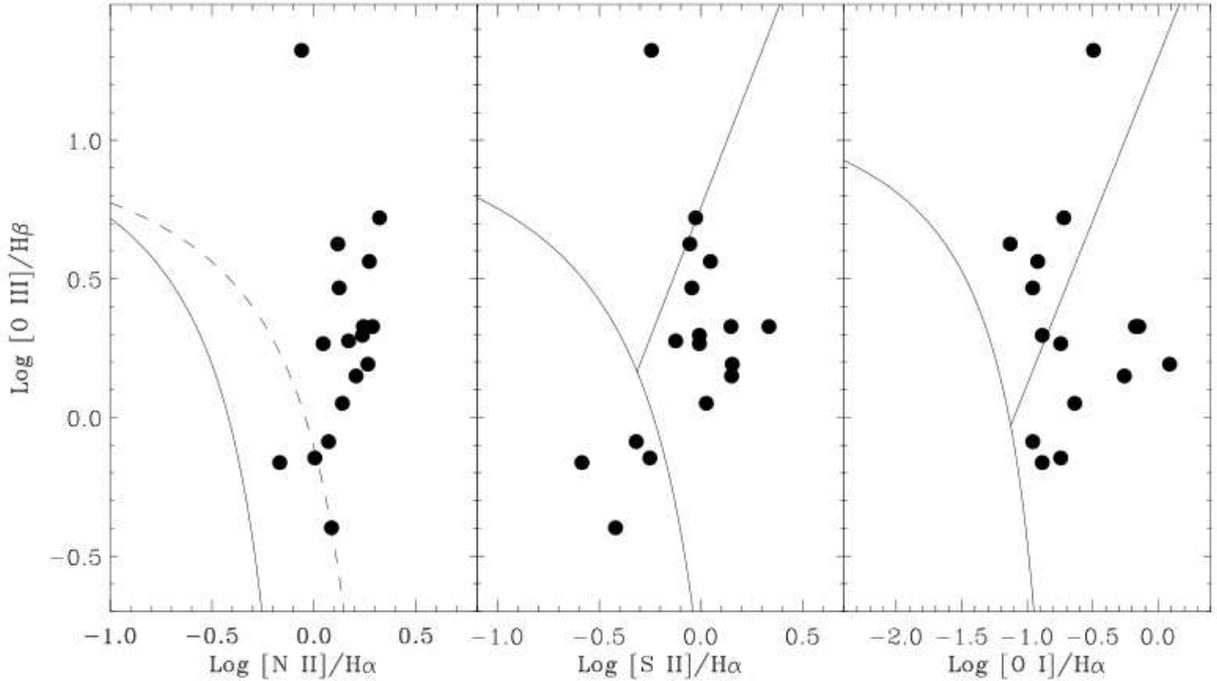}
\caption{Spectroscopic diagnostic diagrams for the 18 galaxies of the SUNNS
  sample. The solid lines are from \citet{kewley06} and separate
  star-forming galaxies, LINER, and Seyfert; in the first panel the region
  between the two curves is populated by the composite galaxies.}
\label{diag}
\end{figure*}

\begin{figure}
\centering
\includegraphics[width=7cm]{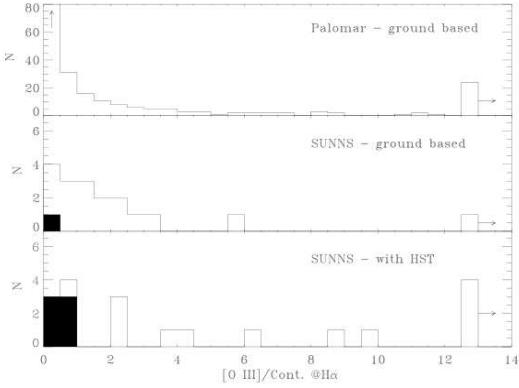}
\caption{\ew\ distribution for the Palomar sample (top panel). The first bin
  contains 299 objects, including 96 emission galaxies whose \oiii\ line
  is not detected.  In the other panels we show instead the \ew\ distribution
  for the SUNNS subsample derived from the Palomar (middle) and the HST
  (bottom) observations. The filled areas indicate \ew\ upper limits; the
  horizontal arrows show the objects with \ew\ larger than 13 \AA.}
\label{istew}
\end{figure}

The sample of 23 galaxies selected for the Survey of Nearby Nuclei with the
Space Telescope Imaging Spectrograph (STIS) on board HST (SUNNS,
\citealt{shields04}) is perfectly suited for our purposes. The SUNNS targets
were drawn from the Palomar survey and include all galaxies within 17 Mpc with
emission lines fluxes greater than $10^{-15}$ $\ergscm$ within a $2\arcsec
\times 4\arcsec$ aperture. An archival search showed, in addition to the HST
spectroscopic data, an almost complete coverage also of Chandra and VLA
observations. The SUNNS galaxies offer a good representation of the various
classes of activity (as derived from the Palomar survey) including LINERs,
Seyfert, and transition galaxies (see Table \ref{table}). We provide  
in Table \ref{table} both the original spectral classification from \citet{ho97}
 and that based on the scheme proposed by \citet{kewley06} into LINERs, Seyfert, and
  ambiguous galaxies\footnote{Ambiguous galaxies change classification. In these cases we report in parenthesis also
    the different classifications. NGC~4596 can be classified as LINERs based
    on its \oiii/\Hb\ lower limit.} derived from their location in the
 three diagnostic diagrams shown in Fig. \ref{diag}. Because we are mainly
interested in active galaxies, we will not consider the 5 star forming
galaxies part of the SUNNS sample. From the point of view of the line
strengths, \ew\ has a distribution similar to that of the whole Palomar sample
(see Fig. \ref{istew}). In particular, the concentration toward low EW values
is still present, although reduced with respect to the whole sample.  This is
due to the selection only of objects of high emission line fluxes.

These objects have been observed with STIS through a 0\farcs2 wide slit with
two grisms (G430L and G750M) that covered the blue and red part of the
optical spectrum with an exposure time of $\sim$1800 and $\sim$3000 s.
 Emission line measurements were obtained by \citet{shields07}
from spectra extracted from a 0\farcs2 $\times$ 0\farcs25 synthetic aperture.

In the Chandra public archive we found data for all but one of the 18 selected
SUNNS galaxies. Observations of most of them are already presented in the
literature, but we preferred, for the sake of homogeneity, to re-analyze these
data.  For each object we extracted the spectrum in a region of 1\farcs5 of
radius, centered on the X-ray source closer to the peak of the optical
emission as measured from HST imaging. The optical/X-ray offsets are all
smaller than $\sim$1\farcs1, as expected from the accuracy of the relative
astrometry between HST and Chandra images. Nevertheless, some
sources are situated in a crowded field where the identification is uncertain (e.g.,
NGC~4321). 

We generally fit the data with a single power-law with the column density
  N$_{\rm H}$ fixed at the Galactic value. In four cases, the observed counts are
  not sufficient to constrain the spectral index $\Gamma$ and we fixed its
  value to 1.7. We explored the possible presence of additional local
  absorption, N$_{\rm H,z}$. The data quality is sufficient to test this
  effect in six galaxies but we do not find any convincing example (based on the
  F-test) of an absorbed source. The corresponding upper limits are in the range
  N$_{\rm H,z} \sim 1-5\times10^{21} {\rm cm}^{-2}$. In two cases, the data
instead show a preference for a more complex model that also includes
thermal emission (i.e., NGC~4203 and NGC~4321).

When no clear point source was present, we estimated a 1 $\sigma$ upper limit,
adopting the model described before and setting the power-law normalization to
reproduce the observed count rate.

All these
sources have high-resolution ($\sim$0\farcs15) radio data available, that were obtained
with the VLA at 15 GHz \citep{nagar05}; five of them are detected in the radio
band (see Table \ref{table}).

\begin{figure*}
\label{xoiii}
\includegraphics[width=8.5cm]{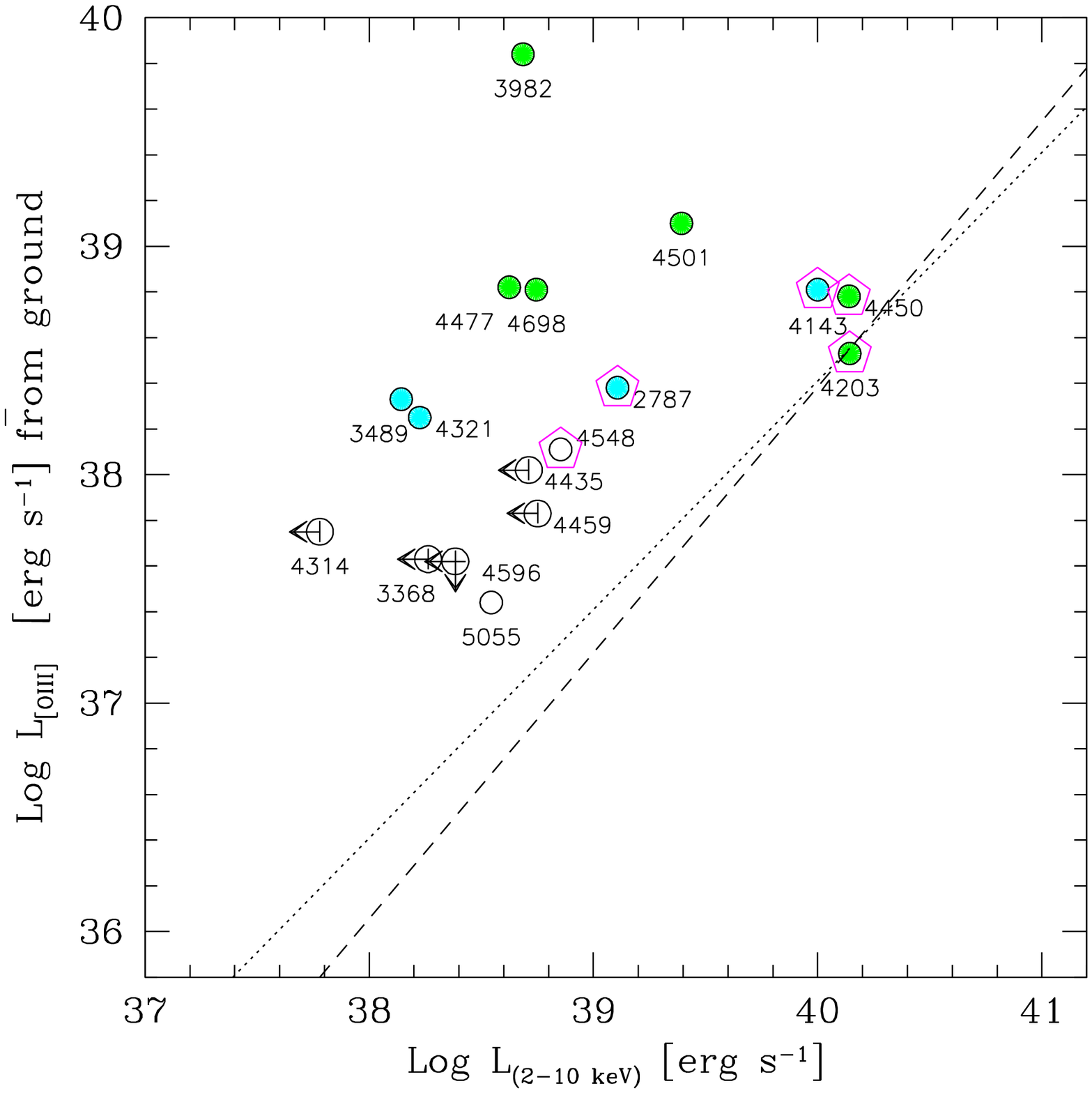}
\includegraphics[width=8.5cm]{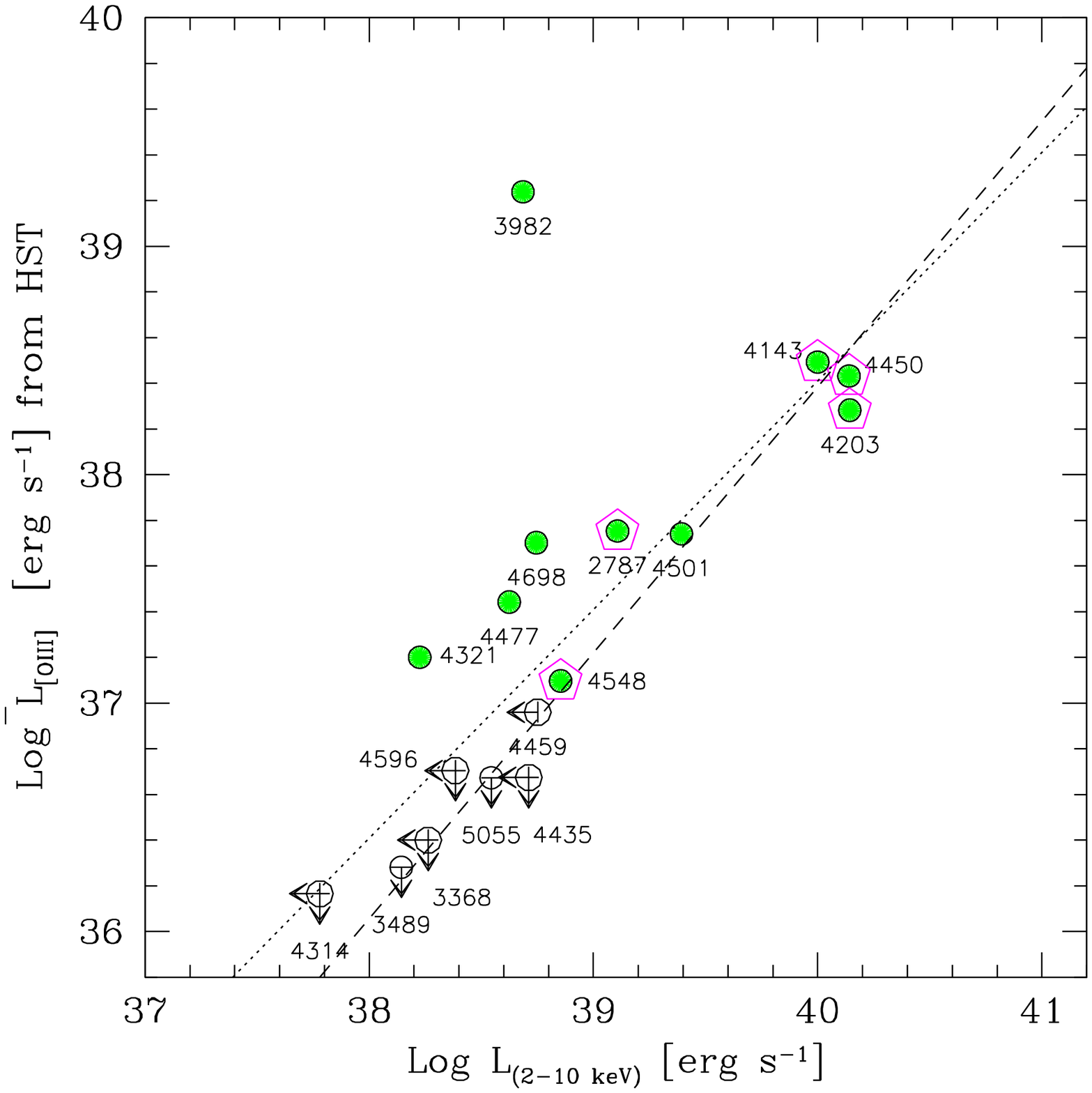}
\caption{\oiii\ versus X-ray luminosity for the SUNNS galaxies. In the left
  panel we use the \oiii\ measurements from the Palomar survey, in the right
  panel we report the STIS data. The symbol colors are coded with the \ew\
  values measured in the relative spectrum: green is for objects with \ew\
  $>$ 2 \AA, cyan for 1 \AA\ $<$ \ew\ $\leqslant$2 \AA, and empty for \ew\
  $\leqslant$ 1 \AA. We denote the sources that are detected in the radio
  with large purple pentagons. The dashed line reproduces the correlation
  between \loiii\ and \lx\ obtained by \citet{panessa06} for a sample of 13
  nearby type I Seyfert galaxies; the dotted line marks the locus of constant
  ratio between line and X-ray luminosities normalized to the value measured
  by \citet{heckman05} in a sample of 20 type I AGNs with bright emission
  lines, i.e., log \lx/\loiii =1.59. The median \oiii\ luminosities of these
  two samples are log \loiii\ $\sim$ 40.0 and $\sim$ 41.6, respectively.}
\end{figure*}

\section{Results}

A strong correlation between the \oiii\ and the X-ray luminosity is known to
exist for unobscured AGN, i.e., type I  Seyfert galaxies and quasars
(e.g., \citealt{heckman05,panessa06}). This can be interpreted as due to the
common origin from the accretion disk of the radiation field
that photoionizes the surrounding gas and produces the line emission on the one hand,
 and of the hard X-ray emission of the disk corona on the other hand.

The SUNNS galaxies do not obey to this general trend (see Fig. \ref{xoiii},
left panel). Only a weak link between \loiii\ and \lx\ is found and most
objects lie well above the relation defined by the more powerful 
  unobscured AGN. Several interpretations can be suggested, for instance
that the spectral energy distribution of these low-luminosity AGN differ from
that of their brighter counterparts, or, alternatively, that they are
  strongly absorbed in the X-ray. However, as reported in Sect. \ref{sample},
  in six galaxies we were able to test the presence of X-ray absorption, always
  with a negative result. Two of these sources (namely NGC~4203 and NGC~4450)
  show broad lines in their HST spectra but in the remaining four (NGC~2787,
  NGC~4143, NGC~4698, and NGC~5055) only narrow lines are seen. This result 
  agrees with the finding that low-luminosity AGN (\lx $\lesssim 10^{40}$
  $\ergs$) are usually unobscured in the X-ray (e.g.,
  \citealt{balmaverde06a,zhang09}) and that the lack of optical broad lines is
  an intrinsic property, not related to selective absorption.

The connection between line and X-ray emission becomes much
clearer from the HST emission line fluxes. All objects move toward
significantly lower luminosities (see Fig. \ref{xoiii}, right panel, and Table
\ref{table}). This indicates that the line emission is extended and that a
fraction of it is lost going from the aperture of the Palomar survey to the
HST (which is $\sim$160 times smaller in area).  The objects with the
lowest line luminosity and \ew\ values are more affected. With the sole
exception of NGC~4459, the objects with \ew\ $<$ 1 and \loiii\ $ < 10^{37}
\ergs$ do not have a detectable \oiii\ line and decrease their luminosity by a
factor $>$ 10. Moreover the emission line flux of some objects 
at the high-luminosity end is
 significantly reduced, by a factor of up to 25, but in all
these sources \ew\ increases.

But the most interesting finding is that with the HST line measurements
a clear trend between \loiii\ and \lx\ emerges with most objects now lying
close to the powerful AGN correlation. Apparently, the small HST aperture
enabled us to isolate the compact region of line emission in the immediate
proximity of the galaxy nucleus, removing the contamination of the diffuse
line emission produced in the host galaxy. This nuclear component is closely
related to the X-ray emission.

A clear exception is NGC~3982: its offset from the correlation remains very
large even considering the HST measurements. \citet{akylas09} showed, with
XMM observations that it is a highly absorbed source (${\rm N_{\rm H}} \sim 4.3
\times 10^{23}$cm$^{-2}$) with an X-ray luminosity ($0.6 \times 10^{40}
\ergs$) much higher than found in our analysis. In the lower signal-to-noise
Chandra spectrum the un-absorbed hard X-ray tail (at energies $\gtrsim$ 3 keV)
seen in the XMM data is not visible, and consequently the X-ray flux is
underestimated. This is generally expected to be the case for highly X-ray
obscured AGN, and indeed Seyfert 2 are all located well above the \loiii\ -
\lx\ correlation \citep{heckman05}. However, these objects can be easily
recognized as active galaxies based on the high value of \ew.

\medskip The novelty of our approach is, in addition to using multi-band data,
that we include the value of \ew\ in the analysis. We now consider
the SUNNS galaxies by separating them on the basis of their EW in the ground-based 
spectra.

In all six  galaxies with \ew\ $>$ 2 \AA\ in the Palomar spectra (see
  Table \ref{table}) the detection in the X-ray (and, in three cases, also in the
  radio band) supports their identification as AGN.  Their \ew\ increases when
using the smaller HST aperture, as expected when the line emission is
spatially concentrated toward the galaxy's nucleus. However, in three of them, the
\oiii\ flux decreases substantially, by a factor of 10 - 25. This is 
expected since HST imaging of nearby AGN {\citep{capetti96,falcke98} usually
  reveals line emission extended on a scale much larger than the region
  covered by the STIS aperture.

  Of the five galaxies with intermediate line strengths, 1 \AA\ $<$ \ew\
  $\leq$ 2 \AA, four show an increased \ew\ in the HST data, reaching values
  between 2.4 \AA\ and 25 \AA. They also show an X-ray nucleus
  (except in the case of NGC~3992 for which there are no Chandra data
  available); two of them are also detected in the radio band. The nature of
  the last galaxy, NGC~3489, is less secure. Although detected in the X-ray,
  there are no emission lines in its HST spectrum and it does not produce
  radio emission at a detectable level. A significant problem linked to this galaxy
  and, more in general to low-luminosity AGN, is the possibility of a
  contamination from stellar sources, e.g., from X-ray binaries. As discussed
  by, e.g., \citet{zhang09}, the probability of such a confusion is relatively
  low, but it cannot be ruled out in individual cases. This problem is
  particularly relevant because of the lack of nuclear optical emission
  lines. Nonetheless, the location of this source in Fig. \ref{xoiii} is
  consistent with a low-luminosity active nucleus. Deeper optical spectroscopic
  data are needed to clarify the nature of this source.

  Finally, there are seven galaxies with \ew\ $\leq$ 1 \AA. Four of them do not
  show any evidence for an active nucleus because there are no visible emission
  lines in their HST spectra and they are not detected in either the X-ray or
  the radio bands. To this list we can safely add NGC~4459 whose \ew\ from
  HST is only 0.6 \AA\ ,while it lacks of X-ray and radio emission.

  But among the objects with a low \ew\ there is also an object with the
    characteristics of an AGN, namely NGC~4548, with an \ew\ growing from 0.9
    \AA\ to 2.5 \AA\ according to the Palomar and HST data, and it is
    detected in both the X-ray and radio data. Finally, NGC~5055 is detected
  in the X-ray, but without emission lines or radio emission, and, similarly
  to NGC~3489, its classification remains uncertain.

\section{Discussion}

We showed that with optical spectroscopy and X-ray imaging at high
spatial resolution, such as can be obtained with HST and Chandra,
our understanding of the nature of emission line galaxies can be
  significantly improved with respect to the results derived from ground-based
  data obtained over large ($\sim$10 square arcsec) apertures, in particular
  concerning the question which among them are AGN and which are not.

  By considering together the \ew\ measured from HST spectroscopy and the
  detection of a nuclear source in the Chandra data we found that the galaxies
  of our sample can be separated into two populations. 

  The 11 objects belonging to the first class have an equivalent width of the
  \oiii\ emission line measured from HST data \ew\ $\gtrsim$ 2 \AA\ and are
  associated with an X-ray nuclear source. They are also characterized by an
  increase of their \ew\ when moving from the ground-based to the HST
  measurements. This indicates that the emission line is more highly
  nuclear concentrated than the continuum (mostly stellar) emission. These
  are most likely active galaxies. 

  In the second group we found seven galaxies with \ew $\lesssim$ 1 \AA\ that
  do not generally show any emission related to an active nucleus (emission
  lines, X-ray, or radio sources). This group includes about half of the
  LINERs (or transition galaxies) of the sample.  These are objects of low
  \oiii\ line luminosity ($\lesssim 10^{38} \ergs$) and low equivalent width
  ($\lesssim 1 $\AA) in ground-based observations.

  The HST spectroscopic data alone are already able to distinguish very
  efficiently between active and non-active galaxies. Indeed, the two classes
  described above are separated into those lacking  nuclear emission lines
  (corresponding to upper limits of $\lesssim$ 1 \AA) and those with \ew\ $>$
  2 \AA. All objects with a low \ew\ in the HST data, and generally also
  those that lack an X-ray central source, are galaxies of low equivalent width in
  the Palomar spectra. This strengthens the suggestion presented in
  \citet{capetti11c} that this parameter is a good predictor of the nature of
  an emission line galaxy. Nonetheless, several objects of low \ew\ ($\sim$ 1
  - 2 \AA) from the ground experience a strong increase when observed at high
  spatial resolution.  Furthermore, the concentration at low \ew\ of the
  ground-based measurements effectively disappears in the HST data; as
  explained in the introduction, this effect is difficult to account for lines
  powered by an AGN.

Similarly powerful is the classification of AGN based on the presence of a
nuclear X-ray source in the Chandra data. The comparison of the results
obtained with the two instruments generally agree with,
confirming the validity of their respective classifications. Only a few
objects may be different. These galaxies
underline the potential limitations of the X-ray data due to i) a high level
of X-ray absorption (as for NGC~3982) or ii) the possible confusion
with stellar sources (for NGC~3489 and NGC~5055).

How do these results affect our understanding of how many 
  emission line galaxies are AGN? Apparently, five (but possibly seven) galaxies out
  of 18 lack of any evidence of an active nucleus. However,
the sample considered is biased by the selection of galaxies of high emission
line fluxes, and indeed their \ew\ distribution is not representative of 
the whole Palomar sample (see Fig. \ref{istew}). In this sample, the
objects located in the AGN region based on their emission line ratios are
$\sim$170 (or $\sim$250 including the composite galaxies); 50\% of them have
\ew\ $<$ 1 \AA\ and 20\% have $1 \AA\ \leq \rm{EW[O~III]} < 2
\AA$. Extrapolating our results, the fraction of non-active galaxies among them
is $\sim$35-45\%.  But, even more importantly, we cannot confirm 
 an active nucleus in any of the galaxies with \loiii\ $\lesssim 10^{38}
\ergs$ as measured from ground-based observations.

\section{Summary and conclusions}

  We explored the nature of a sample of 18 emission line
  galaxies by combining high-resolution observations obtained in different
  bands, with HST, Chandra, and the VLA. All these sources are located in the
  region populated by AGN in the optical spectroscopic diagnostic
  diagrams. However, it is still unclear which of these objects, in particular
  among the LINERs, are powered by an active nucleus and for which objects
  other sources of gas ionization are responsible for the observed emission
  lines.

  No connection was found between X-ray and emission line luminosities from
  ground-based data, unlike what was found for brighter AGN. Conversely, when
  using the HST spectroscopic data, which are extracted on a much smaller aperture, a
  strong correlation emerges, consistent with the link found in more powerful
  unobscured AGN. This indicates that the ground-based line measurements are
  severely affected by the contamination of diffuse emission from the host
  galaxies. Conversely, the small HST aperture enabled us to isolate the
  compact region of line emission in the immediate proximity of the active
  nucleus, when this is present.

  The galaxies separate into two populations. The 11 objects belonging to the
  first class have an equivalent width of the \oiii\ emission line measured
  from HST data \ew\ $\gtrsim$ 2 \AA\ and are associated with an X-ray nuclear
  source; in the second group we find seven galaxies with \ew $\lesssim$ 1 \AA\
  that do not generally show any emission related to an active nucleus
  (emission lines, X-ray, or radio sources). This latter group includes about
  half of the LINERs (or transition galaxies) of the sample, all of which are objects
  of low \oiii\ line luminosity ($\lesssim 10^{38} \ergs$) and low equivalent
  width ($\lesssim 1 $\AA) in ground-based observations. These two groups can
  be associated with a different origin of their emission lines, the first
  being powered by an active nucleus, the second in which other processes lead
  to the formation of the lines. This strengthens the suggestion that the \ew\
  value is a better predictor of the nature of an emission line galaxy than
  its location in the diagnostic diagrams.

How can we further improve the AGN census?  For some of the SUNNS objects, the
main limiting factor is the quality of their HST and Chandra data. While for a
few objects (e.g., NGC~3368 and NGC~4435) significantly deeper Chandra
observations can be obtained with a ``reasonable'' exposure time ($\sim$ a few
tens of ksec), this is generally not the case. Similarly, the HST spectra
already correspond to about one HST orbit and no order of magnitude
improvement can be predicted. From deeper radio data we can expect to obtain
additional confirmation of the AGN presence. However, the ratio between radio and
optical (or X-ray) luminosities for the five radio detected AGN spans a broad
range of $\sim$2 dex. Furthermore, the four galaxies with the highest Palomar
line luminosity are not detected by the VLA observations. This stresses the
difficulty of predicting a level of radio emission for active nuclei. Radio
observation cannot then be used to falsify an AGN, but only to confirm it.

Analyzing a larger number of galaxies is apparently the best way to set
our results on a stronger statistical basis. The HST and Chandra archives already
contain data for a substantial fraction of galaxies of the Palomar galaxies
($\sim$1/4 with STIS/HST and $\sim$2/3 with Chandra). 

A promising alternative approach is to obtain ground-based optical
spectroscopy, but extracted over a relatively small aperture, $\lesssim$1
square arcsec. This produces a better match of the size of the aperture
and of the emission line region, including, with respect to the HST aperture,
a larger part of line emission from any active nucleus, while still
limiting the host contamination. Furthermore, the higher flux included in the
aperture, combined with a more extended telescope collecting area, will substantially 
improve the quality of the spectra. This might allow us to reach lower
levels of \ew, closer to the limit of $\sim$0.1 \AA\ set by the current
limitations in describing galaxy spectra with stellar templates
\citep{sarzi06}. Ideally, integral field units should be used to
also measure the spatial distribution of emission lines and stellar
continuum.

\end{document}